\documentclass[12pt]{article}
\usepackage[english]{babel}
\usepackage{amsmath,amsfonts,amssymb,xspace,url}
\usepackage{graphicx}

\usepackage[
    a4paper,
    textwidth=16cm,
    textheight=26cm,
    centering
]{geometry}

\newcommand{\BelT}{{\scshape BelT}\xspace}
\makeatletter
\renewcommand{\@biblabel}[1]{#1.}
\makeatother

\begin{document}
\begin{sloppypar}

\begin{center}
{\Large\bfseries BelT Block Cipher%
\renewcommand{\thefootnote}{\fnsymbol{footnote}}%
\footnotemark[1]}

\vspace{1.5em}
{\large S.~V.~Agievich$^1$,
V.~A.~Galinskij$^1$,
N.~D.~Mikulich$^2$,
Yu.~S.~Kharin$^1$}

\vspace{1em}
Translated from Russian by N.~I.~Kalosha$^3$

\vspace{1.5em}
\begin{tabular}{c}
$^1$National Research Center for Applied Problems of Mathematics and Informatics\\
Belarusian State University\\
Fr.~Skaryna av.~4, Minsk, Belarus \\[0.6em]
$^2$National Center of Information Security of the Republic of Belarus\\
Kalvariyskaya 7a, Minsk, Belarus \\[0.6em]
$^3$Institute of Mathematics of the National Academy of Sciences of Belarus\\
Surganova 11, Minsk, Belarus
\end{tabular}
\end{center}

\renewcommand{\thefootnote}{\fnsymbol{footnote}}
\footnotetext[1]{Original paper in Russian: \textit{Agievich~S.~V.}, 
\textit{Galinsky~V.~A.}, \textit{Mikulich~N.~D.}, \textit{Kharin~Yu.~S.}, Block 
encryption algorithm BelT.~--- Upravlenie Zaschitoj Informacii, 2002, v.~6, 
No.~4, p.~407--412. Available at: 
\url{https://elib.bsu.by/bitstream/123456789/24140/1/BelT.pdf}.} 
\renewcommand{\thefootnote}{\arabic{footnote}}

\begin{abstract}
BelT is a block cipher with a 128-bit block size and a 256-bit key, developed in
Belarus and adopted as part of the Belarusian national standard STB 34.101.31
for encryption and integrity control. Its specification was issued as a
preliminary standard in 2007, becoming a final standard in 2011. BelT employs an
eight-round structure combining elements of a Feistel network and the Lai–Massey
scheme. This paper presents a complete description of the cipher and explains
the principles underlying its design. Software performance and possible
optimizations of BelT are discussed in comparison with those of candidates in
the Advanced Encryption Standard (AES) competition.
\end{abstract}

\section{Introduction}

Numerous block ciphers have been proposed and standardized for national,
international, and sector-specific applications.  The best-known cryptographic
ciphers include \texttt{DES}, \texttt{GOST~28147-89}, \texttt{IDEA},
\texttt{Rijndael}, and others. They have received considerable attention from
cryptographic researchers, who study their weaknesses and assess the security of
these ciphers.

This paper describes a block cipher called BelT with a 128-bit block size and a
256-bit key. The design principles of the cipher and its performance
characteristics are presented.

\section{Specification}\label{bALGO}

\subsection{Notation}\label{bALGO.Defs}

\begin{tabular}{crp{13.5cm}}
${\cal A}^n$            & \hspace{0mm} &
the set of all binary strings of length $n$ over the 
alphabet~${\cal A}=\{0,1\}$;\\[4pt] 
$u\parallel v$          &&
the concatenation
$u_1 u_2\ldots u_n v_1 v_2\ldots v_m$
of the binary strings
$u=u_1 u_2\ldots u_n$ and
$v=v_1 v_2\ldots v_m$;\\[4pt]
$\bar u$                &&
a)~the integer
$2^7 u_1+2^6 u_2+\ldots+u_8$
if the binary string $u$ has the form
$u=u_1 u_2\ldots u_8\in{\cal A}^8$ and\\[2pt]
                        &&
b)~the integer
$\bar u_1+2^8\bar u_2+2^{16}\bar u_3+2^{24}\bar u_4$
if the binary string $u$ has the form
$u=u_1\parallel u_2\parallel u_3\parallel u_4$, $u_i\in{\cal A}^8$;\\[4pt]
$\langle U\rangle_{32}$ &&
the binary string $u\in{\cal A}^{32}$ such that $\bar u\equiv 
U\pmod{2^{32}}$;\\[4pt]
$u\oplus v$             &&
the bitwise sum modulo 2 of the binary strings $u,v\in{\cal A}^{32}$;\\[4pt]
$u\boxplus v$           &&
the binary string $\langle\bar u+\bar v\rangle_{32}$, where
$u,v$ are 32-bit strings, $u,v\in{\cal A}^{32}$;\\[4pt]
$u\boxminus v$          &&
the binary string $w\in{\cal A}^{32}$ such that $u=v\boxplus w$;\\[4pt]
$\lambda(u)$            &&
the binary string
$\left\langle 2\bar u+\left\lfloor
\frac{\bar u}{2^{31}}\right\rfloor\right\rangle_{32}$ 
where $u$ is a 32-bit binary string, $u\in{\cal A}^{32}$;\\[4pt]
$\lambda^r(u)$          &&
the binary string obtained by applying $\lambda$ to $u$ $r$ times;\\[4pt]
$a\leftarrow u$         &&
assignment of the value $u$ to the register $a$;\\[4pt]
$a\leftrightarrow b$    &&
swapping the contents of registers $a$ and $b$.
\end{tabular}

\subsection{Encryption and decryption}\label{bALGO.Description}

\begin{description}

\item{{\bf Input data.}}
The cipher is intended for encryption and decryption of blocks 
$X\in{\cal A}^{128}$ under keys $\theta\in{\cal A}^{256}$.

\item{{\bf Key schedule.}}
A key $\theta=\theta_1\parallel\theta_2\parallel\ldots\parallel\theta_8$, 
$\theta_i\in{\cal A}^{32}$, is used to generate the round keys 
$\kappa_1,\kappa_2,\ldots,\kappa_{56}\in{\cal A}^{32}$. 
For encryption, $\kappa_i$ is the $i$-th element of the sequence
$$
\theta_1,\theta_2,\ldots,\theta_8,\theta_1,\theta_2,\ldots,\theta_8,\ldots
$$
For decryption, $\kappa_i$ is the $i$-th element of the sequence
$$
\theta_8,\theta_7,\ldots,\theta_1,\theta_8,\theta_7,\ldots,\theta_1,\ldots
$$

\item{{\bf Cipher structure.}}
A block $X_1\parallel X_2\parallel X_3\parallel X_4$, $X_i\in{\cal A}^{32}$,
is processed using 32-bit registers $a,b,c,d$, initialized as
$$
a\leftarrow X_1,\quad
b\leftarrow X_2,\quad
c\leftarrow X_3,\quad
d\leftarrow X_4.
$$
Encryption and decryption are defined as operations on these registers,
which are grouped into rounds. An additional 32-bit auxiliary register $e$
is also used.

For encryption, rounds $1,2,\ldots,8$ are performed. The ciphertext is the
block $b\parallel d\parallel a\parallel c$.

For decryption, rounds $8,7,\ldots,1$ are performed. The plaintext is the
block $c\parallel a\parallel d\parallel b$.

\item{{\bf Rounds.}}
During encryption round $t$, the following steps are performed
(see Fig.~\ref{Figure.bALGO.Round}):
$$
\begin{array}{rl}
(1) & b\leftarrow b\oplus    G_{5}(a\boxplus\kappa_{7t-6}),\\
(2) & c\leftarrow c\oplus    G_{21}(d\boxplus\kappa_{7t-5}),\\
\\
(3) & a\leftarrow a\boxminus G_{13}(b\boxplus\kappa_{7t-4}),\\
\\
(4) & e\leftarrow G_{21}(b\boxplus c\boxplus\kappa_{7t-3})\oplus
\langle t\rangle_{32},\\
(5) & b\leftarrow b\boxplus  e,\\
(6) & c\leftarrow c\boxminus e,\\
\\
(7) & d\leftarrow d\boxplus  G_{13}(c\boxplus\kappa_{7t-2}),\\
\\
(8) & b\leftarrow b\oplus    G_{21}(a\boxplus\kappa_{7t-1}),\\
(9) & c\leftarrow c\oplus    G_{5}(d\boxplus\kappa_{7t}),\\
\\
(10) & a\leftrightarrow b,\\
(11) & c\leftrightarrow d,\\
(12) & b\leftrightarrow c.
\end{array}
$$

\begin{figure}[bht]
\begin{center}
\includegraphics[width=12cm]{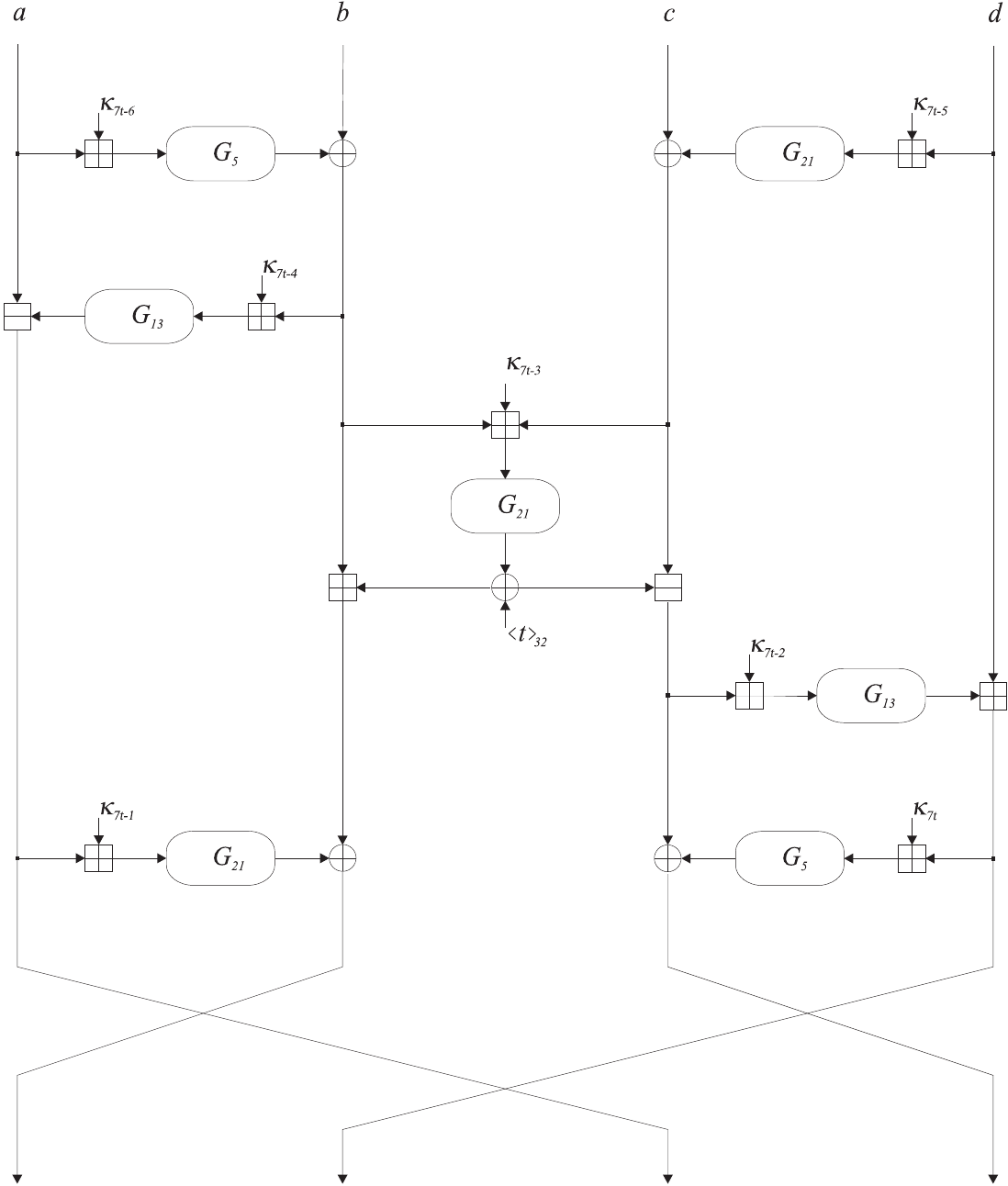}
\end{center}
\caption{Encryption round $t$}\label{Figure.bALGO.Round}
\end{figure}

During decryption round $t$, step~(12) is replaced by
$$
\begin{array}{rl}
(12') & a\leftrightarrow d.\\
& \phantom{e\leftarrow G_{15}(b\boxplus c\boxplus\kappa_{7t-3})\oplus
\langle t\rangle_{32},}
\end{array}
$$

\item{{\bf Transformation $G_r$.}}
Transformation $G_r\colon{\cal A}^{32}\to{\cal A}^{32}$
maps the string
$u=u_1\parallel u_2\parallel u_3\parallel u_4$,
$u_i\in{\cal A}^8$, to the string
$$
G_r(u)=\lambda^r\left(
S(u_1)\parallel S(u_2)\parallel
S(u_3)\parallel S(u_4)
\right).
$$

\item{{\bf $S$-box $S$.}}
The $S$-box $S\colon{\cal A}^8\to{\cal A}^8$ is specified by
Table~\ref{Table.bALGO.SBox}. Binary strings are represented in hexadecimal
notation. Four consecutive binary digits correspond to one hexadecimal digit.
For example, $10100010={\rm A2}_{16}$. If $u={\rm IJ}_{16}$, then the value
$S(u)$ is found at the intersection of row ${\rm I}$ and column ${\rm J}$. For
example, $S({\rm A2}_{16})={\rm 9B}_{16}$.
\end{description}

\begin{table}[bht]
\caption{$S$-box $S$}\label{Table.bALGO.SBox}
\begin{center}
{\small\tt
\begin{tabular}{|l|rrrrrrrrrrrrrrrr|}
\hline
 &  0&  1&  2&  3&  4&  5&  6&  7&  8&  9&  A&  B&  C&  D&  E&  F\\
\hline
0& B1& 94& BA& C8& 0A& 08& F5& 3B& 36& 6D& 00& 8E& 58& 4A& 5D& E4\\
1& 85& 04& FA& 9D& 1B& B6& C7& AC& 25& 2E& 72& C2& 02& FD& CE& 0D\\
2& 5B& E3& D6& 12& 17& B9& 61& 81& FE& 67& 86& AD& 71& 6B& 89& 0B\\
3& 5C& B0& C0& FF& 33& C3& 56& B8& 35& C4& 05& AE& D8& E0& 7F& 99\\
4& E1& 2B& DC& 1A& E2& 82& 57& EC& 70& 3F& CC& F0& 95& EE& 8D& F1\\
5& C1& AB& 76& 38& 9F& E6& 78& CA& F7& C6& F8& 60& D5& BB& 9C& 4F\\
6& F3& 3C& 65& 7B& 63& 7C& 30& 6A& DD& 4E& A7& 79& 9E& B2& 3D& 31\\
7& 3E& 98& B5& 6E& 27& D3& BC& CF& 59& 1E& 18& 1F& 4C& 5A& B7& 93\\
8& E9& DE& E7& 2C& 8F& 0C& 0F& A6& 2D& DB& 49& F4& 6F& 73& 96& 47\\
9& 06& 07& 53& 16& ED& 24& 7A& 37& 39& CB& A3& 83& 03& A9& 8B& F6\\
A& 92& BD& 9B& 1C& E5& D1& 41& 01& 54& 45& FB& C9& 5E& 4D& 0E& F2\\
B& 68& 20& 80& AA& 22& 7D& 64& 2F& 26& 87& F9& 34& 90& 40& 55& 11\\
C& BE& 32& 97& 13& 43& FC& 9A& 48& A0& 2A& 88& 5F& 19& 4B& 09& A1\\
D& 7E& CD& A4& D0& 15& 44& AF& 8C& A5& 84& 50& BF& 66& D2& E8& 8A\\
E& A2& D7& 46& 52& 42& A8& DF& B3& 69& 74& C5& 51& EB& 23& 29& 21\\
F& D4& EF& D9& B4& 3A& 62& 28& 75& 91& 14& 10& EA& 77& 6C& DA& 1D\\
\hline
\end{tabular}
} 
\end{center}
\end{table}

\subsection{Test Vectors}\label{bDUMP}

Encryption of the string
\begin{align*}
&\mbox{{\tt 504C4149 4E544558 54313642 59544553}}_{16}\
\intertext{under the key}
&\mbox{{\tt 454E4352 59505449 4F4E2F44 45435259
5054494F 4E204B45 59333242 59544553}}_{16}\
\intertext{produces the string}
&\mbox{{\tt BD7611FF 2EC7545C 0734381E 30CBB2DD}}_{16}.
\end{align*}

Table~\ref{Table.bDUMP.Encrypt} gives the contents of registers $a$,
$b$, $c$, and $d$ during the encryption rounds.

\begin{table}[htb]
\caption{Register contents after round $t$}
\label{Table.bDUMP.Encrypt}
\begin{center}
\begin{tabular}{|c|c|c|c|c|}
\hline
Round number  & \multicolumn{4}{|c|}{Registers}\\
 \cline{2-5}
$t$  &$a$&$b$&$c$&$d$ \\
\hline
\hline
$1$&${\tt B4B0505E}_{16}$&${\tt A8569AD5}_{16}$&${\tt FCD5D3F6}_{16}$&${\tt 00FD3257}_{16}$\\
\hline
$2$&${\tt 6C53A06D}_{16}$&${\tt 23666D8C}_{16}$&${\tt D36D40F1}_{16}$&${\tt 29A9BB95}_{16}$\\
\hline
$3$&${\tt 70BBEDCC}_{16}$&${\tt 636AD2FB}_{16}$&${\tt CBFD6800}_{16}$&${\tt 98A20C92}_{16}$\\
\hline
$4$&${\tt 2DD1854F}_{16}$&${\tt 9597DE3A}_{16}$&${\tt B14A5564}_{16}$&${\tt 3698CA8A}_{16}$\\
\hline
$5$&${\tt EDEBACCB}_{16}$&${\tt A7CF3B6B}_{16}$&${\tt 3530E6F4}_{16}$&${\tt F7E87E44}_{16}$\\
\hline
$6$&${\tt 7A15B4B0}_{16}$&${\tt F8B346CD}_{16}$&${\tt 786216BA}_{16}$&${\tt 6FDB2149}_{16}$\\
\hline
$7$&${\tt 334B8604}_{16}$&${\tt E4A8697B}_{16}$&${\tt 937FF79A}_{16}$&${\tt FEC1A580}_{16}$\\
\hline
\end{tabular}
\end{center}
\end{table}

\section{Design Principles}\label{bDESIGN}

The following main principles guided the design of the cipher.

\begin{enumerate}
\item
Structural similarity of the encryption and decryption transformations.

The BelT encryption round differs only slightly in structure from the
decryption round. This similarity enables simpler software and hardware
implementations of the cipher.

\item
Customization of the $S$-box $S$.

If a personalized cipher is required, the cipher can use a custom 
$S$-box $S$ as an additional long-term key.

\item
Construction and properties of the $S$-box $S$.

The truth tables of the coordinate Boolean functions of $S$ were
selected as distinct length-$255$ segments of linear recurrence sequences
(see~\cite{LidlNiederreiter97}) with the primitive characteristic polynomial
$$
p(\lambda)=
\lambda^8+\lambda^6+\lambda^5+\lambda^2+1,
$$
with the symbol $0$ inserted at a fixed position in each segment.

Some cryptographic properties of the resulting $S$-box are as follows:
\begin{itemize}
\item[1)]
The nonlinearity of $S$ is 102
(for the selected dimensions, the nonlinearity of $S$ cannot exceed 120;
high nonlinearity provides resistance to linear cryptanalysis~\cite{Mat93});

\item[2)]
The differential properties of $S$ are characterized by the quantities
$R_{\oplus\oplus}(S)=8$,
$R_{\boxplus\boxplus}(S)=7$,
$R_{\oplus\boxplus}(S)=6$,
$R_{\boxplus\oplus}(S)=3$,
where, for example,
$$
R_{\boxplus\oplus}(S)=
\max_{
{\scriptstyle\alpha,\beta\in{\cal A}^8
\atop
\scriptstyle\alpha\neq 0}
}
\sum_{x\in{\cal A}^8}{\rm I}\{S(x\boxplus\alpha)=S(x)\oplus\beta\},
$$
and ${\rm I}\{{\cal E}\}$ denotes the indicator function of the event 
${\cal E}$. These values are sufficiently small to provide resistance to 
differential cryptanalysis~\cite{BihSha91};

\item[3)]
The algebraic degree of every coordinate Boolean function of $S$ is 7,
which is the maximum possible value for the selected dimensions.
This hinders higher-order differential attacks~\cite{Knu95};

\item[4)]
There are no quadratic relations connecting the inputs and outputs of the
$S$-box. Existence of such relations simplifies algebraic
attacks~\cite{CouPie02}.
\end{itemize}

\item
Alternation of operations.

As encryption/decryption algorithms are executed, the contents of the registers
are combined with outputs of $G$ using the operations $\oplus$, $\boxplus$, and
$\boxminus$. Bitwise XOR is interleaved with addition or subtraction modulo
$2^{32}$. For example, during encryption, the register $b$ is updated as
follows:
\begin{itemize}
\item[($\oplus$)]
 $b\leftarrow b\oplus G_5(a\boxplus\kappa_1)$,

\item[($\boxplus$)]
$b\leftarrow b\boxplus  
\left(G_{21}(b\boxplus c\boxplus\kappa_4)\oplus\langle 1\rangle_{32}\right)$, 

\item[($\oplus$)]
 $b\leftarrow b\oplus G_{21}(a\boxplus\kappa_6)$.
\end{itemize}
The next update (after swapping the registers $a$ and $b$) is
\begin{itemize}
\item[($\boxminus$)]
$a\leftarrow a\boxminus G_{13}(b\boxplus\kappa_{10})$,
\end{itemize}
and so on.
The use of operations from different algebraic groups and their alternation
makes certain cryptanalytic methods more difficult to apply.

\item
Non-uniform rounds.

During step~(4) of each round, the contents of the register $e$ are combined
with the round number $\langle t\rangle_{32}$, and all rounds are distinct
permutations even when the same round keys $\kappa_{7t-6},\ldots,\kappa_{7t}$
are used.
This reduces the chances that the encryption permutations $F_\theta$,
$\theta\in{\cal A}^{256}$, exhibit algebraic weaknesses such as
\begin{itemize}
\item[--]
low order of permutations $F_\theta$---in particular, the existence of weak 
keys $\theta_w$ such that $F_{\theta_w}^2$ is the identity permutation;

\item[--]
equivalent keys, i.e., distinct keys $\theta_1,\theta_2$ such that
$F_{\theta_1}=F_{\theta_2}$.
\end{itemize}

\item
The number of rounds.

The choice of eight rounds was motivated by the need to protect against common
cryptanalytic methods and to establish a comfortable security margin.
During the development of the cipher, preliminary statistical analysis was 
performed to detect statistical dependencies between plaintext, ciphertext,
and key.
It was found that significant statistical regularities are absent already for
a 2-round construction.
\end{enumerate}

\section{Performance}\label{bSPEED}

A 32-bit architecture was chosen as the basis for the cipher's design.
The only operation that is not 32-bit is byte substitution through the $S$-box
$S$.
For the software implementation of the encryption and decryption algorithms, 
the following optimizations were used:
\begin{itemize}
\item[(a)]
For $r=5, 13, 21, 29$, expanded tables specifying the mappings
$$
S_r\colon{\cal A}^8\to{\cal A}^{32},\quad
x\mapsto \lambda^r(S(x)\parallel 0^{24}),
$$
are precomputed, where $0^{24}$ denotes the binary string of 24 zeros.
Storing all four tables requires 4~KB of memory.

\item[(b)]
The value of $G_5(x)$,
$x=x_1\parallel x_2\parallel x_3\parallel x_4$,
$x_i\in{\cal A}^8$, is computed as
$$
G_5(x)=
S_5(x_1)\oplus S_{13}(x_2)\oplus S_{21}(x_3)\oplus S_{29}(x_4).
$$
Similarly,
\begin{gather*}
G_{13}(x)=S_{13}(x_1)\oplus S_{21}(x_2)\oplus S_{29}(x_3)\oplus S_{5}(x_4),\\
G_{21}(x)=S_{21}(x_1)\oplus S_{29}(x_2)\oplus S_{5}(x_3)\oplus S_{13}(x_4).
\end{gather*}
\end{itemize}

The number of operations required for encryption (decryption) of a 16-byte
data block is given in Table~\ref{Table.bSPEED.Operation}.

\begin{table}[htb]
\caption{Number of operations for encryption/decryption of a 16-byte
block}\label{Table.bSPEED.Operation}
\begin{center}
\begin{tabular}{|l|r|}
\hline
Operations                   & Count\\
\hline
\hline
$\boxplus$, $\boxminus$    & 96\\
$\oplus$                   & 40\\
$\lambda^r$                & 56\\
$S$                        & 224\\
$\leftrightarrow$          & 23 \\
\hline
Total                      & 439\\
\hline
\end{tabular}
\end{center}
\end{table}

The performance of BelT was compared with that of the block ciphers from the 
AES competition---candidates for the U.S. encryption standard~\cite{AES}. We
selected the ciphers \texttt{RC6}, \texttt{Rijndael}, \texttt{MARS},
\texttt{Twofish}, \texttt{Serpent}, \texttt{DFC}, and \texttt{Safer+}. The first
five advanced to the second round of the AES competition, and \texttt{Rijndael}
won the AES competition.

Performance comparison was done using C language software implementations of
these ciphers written by B.~Gladman~\cite{Gladman}. Our implementation of
BelT used the same conventions as these implementations, and did not introduce
any new optimization types. For example, expanded $S$-box tables for computing
$G_r$ described above were also used in the implementation of 
\texttt{Rijndael}. According to unofficial information, the implementation of 
\texttt{Serpent} also used this optimization.

Performance benchmarks were performed using the VTune Performance Analyzer
utility~\cite{VTUNE}.

This utility makes it possible to determine the number of CPU cycles required to
execute a program on Pentium-family processors, taking into account the CPU
microarchitecture (pipelining, instruction cache, data cache, and so on).
Benchmarking results are presented in Table~\ref{Table.bSPEED.Speed}. The table
shows that the cipher achieves an encryption speed of
$200\cdot\frac{128}{1367}\approx 18.73$~Mbit/s on a Pentium-200 CPU,
and $1000\cdot\frac{128}{682}\approx 187.7$~Mbit/s on a Pentium~III-1000 CPU.

\begin{table}[htb]
\caption{Pentium-family CPU cycles required to encrypt 16 bytes of plaintext}
\label{Table.bSPEED.Speed}
\begin{center}
{\small
\begin{tabular}{|l|c|c|c|c|c|c|c|c|}
\hline
Processor  & \multicolumn{7}{|c|}{AES candidates}&\\
\cline{2-8}&
\texttt{RC6} &
\texttt{Rijndael} &
\texttt{MARS} &
\texttt{Twofish} &
\texttt{Serpent} &
\texttt{DFC} &
\texttt{Safer+}&
\BelT\\
\hline
\hline
\verb|Pentium|         &
920&  1022&   954&   892&  1693&  3750&  3789&  1367\\
\verb|Pentium II|      &
271&   399&   388&   404&   964&  1246&  1910&   642\\
\verb|Pentium III|     &
281&   427&   410&   429&   984&  1256&  1928&   682\\
\hline
\end{tabular}
} 
\end{center}
\end{table}

An additional advantage of the cipher is a fast method for determining the
round keys $\kappa_1,\ldots,\kappa_{56}$. In practical implementations, it
doesn't require any calculations.
A complex key schedule can reduce the efficiency of a cipher in
applications that require frequent key changes---for example, certain hash 
function constructions.
For reference, the computational complexity of the \verb|Twofish| key schedule
is approximately five times the complexity of encrypting a 16-byte block.

\end{sloppypar}
\end{document}